**Bi-scale Car-following Model Calibration for Corridor Based on Trajectory**

Keke Long[1], Haotian Shi[1*], Zhiwei Chen[2], Zhaohui Liang[1], Xiaopeng Li[1*], Felipe de Souza[3]

[1] Department of Civil & Environmental Engineering, University of Wisconsin-Madison, Madison, Wisconsin, 53706, USA

[2] Civil, Architectural and Environmental Engineering, Drexel University, Philadelphia, Pennsylvania, 19104, USA

[3] Argonne National Laboratory, 9700 S. Cass Ave., Lemont, IL 60439, USA

Corresponding Authors: Xiaopeng Li, xli2485@wisc.edu; Haotian Shi, hshi84@wisc.edu

**ABSTRACT**

The precise estimation of macroscopic traffic parameters, such as travel time and fuel consumption, is essential for the optimization of traffic management systems. Despite its importance, the comprehensive acquisition of vehicle trajectory data for the calculation of these macroscopic measures presents a challenge. To bridge this gap, this study aims to calibrate car-following models capable of predicting both microscopic measures and macroscopic measures. We conduct a numerical analysis to trace the cumulative process of model prediction errors across various measurements, and our findings indicate that macroscopic measures encapsulate the accumulation of model errors. By incorporating macroscopic measures into vehicle model calibration, we can mitigate the impact of noise on microscopic data measurements. We compare three car-following model calibration methods: MiC (using microscopic measurements), MaC (using macroscopic measurements), and BiC (using both microscopic and macroscopic measurements)—utilizing real-world trajectory data. The BiC method emerges as the most successful in reconstructing vehicle trajectories and accurately estimating travel time and fuel consumption, whereas the MiC method leads to overfitting and inaccurate macro-measurement predictions. This study underscores the importance of bi-scale calibration for precise traffic and energy consumption predictions, laying the groundwork for future research aimed at enhancing traffic management strategies.

# 1 INTRODUCTION

A car-following model is a mathematical representation describing the dynamic positioning of vehicles in a traffic stream, particularly focusing on how each vehicle follows the one ahead. Car-following model calibration is adjusting model parameters to ensure that the simulated behavior of vehicles in a longitudinal direction closely matches observed real-world behavior. In traffic planning and infrastructure development, a calibrated car-following model is essential for understanding vehicle behavior, such as the cruising or acceleration process, which influences the overall flow and safety of traffic. Additionally, when assessing system-level measurements for an intersection or corridor, obtaining real trajectories for all vehicles is challenging. However, a well-calibrated car-following model can effectively simulate these trajectories, thereby facilitating the extraction of important metrics such as delay and fuel consumption (Song et al., 2013). The accuracy and reliability of traffic simulations heavily depend on the precision of the car-following model calibration.

The most common way of car-following model calibration is called microscopic calibration, which involves an objective function that aims to reduce discrepancies between actual and simulated microscopic metrics, such as individual vehicle speed, acceleration, space headway, and time gap. This calibration method, as documented in Table 1, involves analyzing pairs of consecutive vehicles from a trajectory dataset. The most common calibration procedure in this category involves considering the two consecutive vehicles from a trajectory dataset as a car-following pair. Given the status of the previous vehicle, the observed status of the following vehicle, and the simulated status of the following vehicle generated by the car-following model, an optimization problem can be solved to find the best car-following model parameters that minimize the difference between observed and simulated status. This kind of approach deals with vehicle status in independent time steps (Alfa and Neuts, 1995), which fails to consider the memory mechanism of the car-following behavior and may, therefore, result in suboptimal performance. A new microscopic calibration strategy has been proposed (Li et al., 2016), where the car-following model generates the whole trajectory based on the information of the preceding vehicle and only the initial status of the following vehicle, yielding better performance in overall fitting. However, solely relying on microscopic measurements can potentially be sensitive to data noise or disturbances (Punzo et al., 2011). Besides, these models, even when they exhibit a great performance in microscopic measurements, may not accurately produce macroscopic measurements. This is primarily because modeling errors at the microscopic level are easy to accumulate at the macroscopic level over time and space (Song et al., 2015).

To overcome the low predictability in macroscopic measurements of microscopic calibration, some studies adopt macroscopic calibration (Hourdakis et al., 2003; Li et al., 2016; Ma and Qu, 2020; Mo et al., 2021; Papathanasopoulou and Antoniou, 2015). This calibration method adopts the macroscopic measurements of trajectories simulated by car-following models as the calibration metrics. Macroscopic measurements refer to aggregated data over larger spatial/temporal scales. They provide a generalized representation of the traffic system, capturing overall behavior rather than the actions of individual units. In the context of traffic flow and vehicle dynamics, typical examples include average speed, average travel time, traffic density, and average fuel consumption over certain periods and/or spatial extents. Compared with microscopic measurements, macroscopic measurements, over time and space, exhibit reduced sensitivity to such noises and disturbances. They provide a more consolidated and consistent representation of the system's behavior, thereby offering a more stabilized metric for calibrating car-following models. Previous research calibrated a car-following model to reconstruct trajectories that consider various driving modes, consistently using fuel consumption as the macroscopic metric (Song et al., 2015). This approach offered a more persuasive energy consumption estimation compared to microscopically calibrated models. Nevertheless, different driving conditions can result in identical fuel consumption. The extent to which the vehicle behavior depicted by this model accurately represents realistic microscopic behavior needs for further exploration.

Given the drawbacks of both microscopic and macroscopic calibrations, it's important to consider both metrics when calibrating vehicle-following models. While microscopic measurements allow the model to accurately depict vehicle movement, macroscopic measurements are incorporated as regularization terms and could help in addressing the overfitting problem by leveraging broader traffic behavior patterns to provide a more balanced, generalized model. The research proposed a bi-scale calibration method that includes both microscopic and macroscopic measurements, summarized in Table 1.

**TABLE 1 Conclusion of calibration measurements adopted in the previous study.**

| References | Microscopic measurements | | | | | Macroscopic measurements | |
|---|---|---|---|---|---|---|---|
| | Position | Speed | Acceleration | Space headway | Time gap | Mobility | Energy |
| (Ma and Qu, 2020) | √ | | | | | | |
| (Mo et al., 2021) | | | √ | | | | |
| (Hourdakis et al., 2003; Li et al., 2016; | | √ | | | | | |

| Reference | | | | |
|---|---|---|---|---|
| Papathanasopoulou and Antoniou, 2015) | | | | |
| (Treiber and Kesting, 2013) | √ | | √ | |
| (Chen et al., 2010; Punzo and Simonelli, 2005; Shang et al., 2022; Vasconcelos et al., 2014) | | √ | | |
| (Kesting and Treiber, 2008) | | | √ | |
| (Huang et al., 2018; Kesting and Treiber, 2008; Zhu et al., 2018) | √ | √ | | |
| (Punzo et al., 2012) | | | | √ |
| (Song et al., 2015) | | | | √ |
| (Pourabdollah et al., 2018) | √ | √ | | √ |

In summary, car-following model calibration is essential for the accurate estimation of both microscopic behavior and macroscopic measurements. To enhance the predictability of a car-following model, incorporating both measurement types as criteria in the calibration process could result in closer alignment between simulated and actual values. However, very few studies have included energy consumption in modeling vehicle dynamics, and the combination of both measurements has been lacking in the literature. Additionally, there is a scarcity of research analyzing the accumulation of errors in both microscopic and macroscopic indicators, which could be used to guide the selection of calibration parameters.

To compensate for this research gap, this work proposes a bi-scale calibration method that concurrently minimizes both the microscopic non-aggregated vehicular behavior parameters, such as acceleration and the macroscopic aggregated parameters, such as average travel time and average fuel consumption. This approach enables the derived car-following models to better replicate both microscopic behaviors and macroscopic measurements. Simultaneously, the macroscopic measurements can exert a corrective effect on the microscopic level, mitigating the impact of data noise on microscopic calibration. To validate the rationality of incorporating different scales of measurement into the calibration, we initially constructed a theoretical model of error propagation. This model indicates that, for an individual vehicle, inaccuracies in acceleration can compound over time, accumulating more significantly in cumulative measurements such as speed and position. In scenarios involving multiple vehicles, errors further accumulate in measurements such as average travel time and average fuel consumption. Therefore, a car-following model calibration method is proposed that takes into account both microscopic (e.g., individual acceleration) and macroscopic (e.g., average travel time, average fuel consumption) measurements. To validate this model with real-world data, we obtained extensive vehicle trajectory data at the corridor level over a prolonged duration and wide range through drone video analysis. Subsequently, we calibrated the model using corridor-level data, achieving a car-following model with precise reconstruction capabilities for both microscopic motion and macroscopic measurements.

The contributions of this paper can be delineated into three parts: First, we conducted a theoretical analysis, which analyzes the error propagation across microscopic measurements and macroscopic measurements. Furthermore, this method takes into account both the microscopic transient behaviors and the macroscopic information spanning across time and space. As a result, it can adeptly replicate both microscopic and macroscopic measurements. Moreover, by considering macroscopic indicators, the method can circumvent the detrimental effects of data errors on calibration. Thus, the obtained car-following model can reproduce not only microscopic characteristics but also macroscopic characteristics.

The disposition of this paper is as follows. Section 2 conducted a theoretical analysis and proposed the new calibration method. Section 3 explains the calibration experiment using real-world trajectory data and compares the performance of three categories of car-following model calibration methods. Section 4 analyzes the calibration results. Section 5 concludes the paper and discusses future research directions.

## 2 METHODOLOGY

The workflow of this chapter is illustrated in Figure 1. Initially, to investigate the impact of microscopic errors on different metrics, this section conducts a theoretical analysis of error propagation across various measurements, including acceleration, speed, and position. In the scenario of a single vehicle, we explore the effects of acceleration errors on both speed and position. In scenarios involving multiple vehicles, we analyze how the acceleration errors accumulate over time and traffic flow. The results of the theoretical analysis motivated us to propose a bi-scale calibration method.

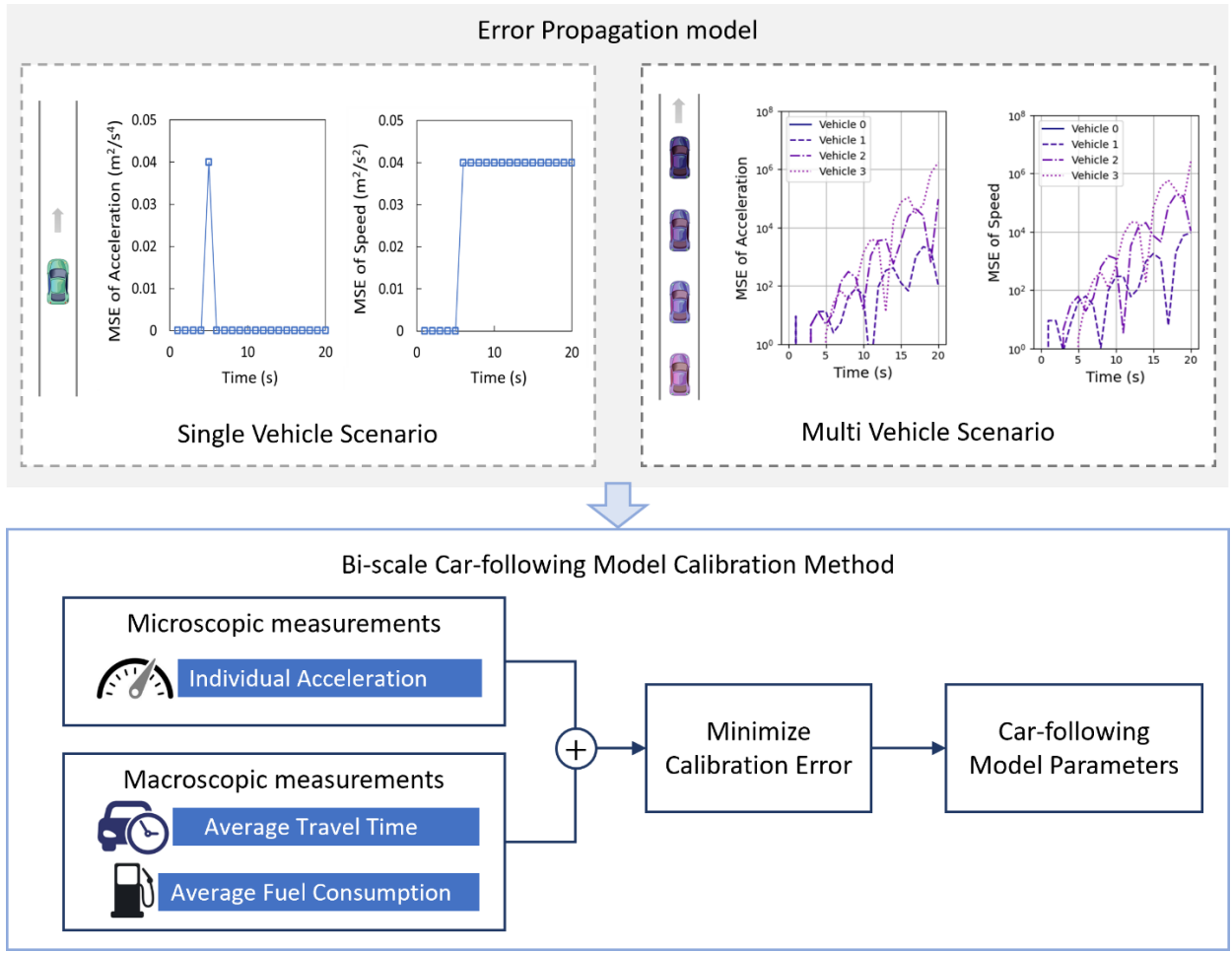

**Figure 1 Methodology flow chart.**

## 2.1 Error Propagation Model

This section aims to theoretically analyze the impact of the acceleration error on both microscopic and macroscopic measurements. We first investigate the single-vehicle scenario, examining the accumulation of acceleration errors in the speed and longitudinal position of a vehicle over time. Second, we extend the analysis to a multi-vehicle scenario using a linear car-following model, revealing how the acceleration error propagates over time and vehicles.

In the theoretical analysis below, we consider a simulation over a time horizon with a set of discrete time points indexed as $t \in \mathcal{T} \coloneqq \{0, 1, \cdots, T\}$ on a single-lane road segment. A group of vehicles indexed as $n \in \mathcal{N} \coloneqq \{0, 1, \cdots, N\}$, is simulated. Denote the simulated position, speed, and acceleration of vehicle $n \in \mathcal{N}$ at time $t \in \mathcal{T}$ as $x_{n,t}^{sim}$, $v_{n,t}^{sim}$, and $a_{n,t}^{sim}$, respectively. Correspondingly, real-world observation of the position, speed, and acceleration of vehicle $n \in \mathcal{N}$ at time $t \in \mathcal{T}$ is denoted as as $x_{n,t}^{obs}$, $v_{n,t}^{obs}$, and $a_{n,t}^{obs}$, respectively. During the simulation, aside from the first vehicle $n = 0$, the accelerations of other vehicles (i.e., $a_{n,t}^{sim}, \forall t \in \mathcal{T}, n \geq 1$) are predicted using a car-following model. The acceleration error is defined as: $\varepsilon_{n,t}^{a} = a_{n,t}^{sim} - a_{n,t}^{obs}, \forall t \in \mathcal{T}, n \in \mathcal{N}$. Similarly, speed errors and position errors are defined as $\varepsilon_{n,t}^{v} = v_{n,t}^{sim} - v_{n,t}^{obs}$ and $\varepsilon_{n,t}^{x} = x_{n,t}^{sim} - x_{n,t}^{obs}, \forall t \in \mathcal{T}, n \in \mathcal{N}$, respectively.

*2.1.1 Error Propagation of Single Vehicle*

We first consider the scenario of a single vehicle $n \in \mathcal{N}$. For the convenience of the notation, we omit the vehicle index $n$ in the remainder of this section. Assume that the errors on all three variables are 0 at the beginning of the simulation, i.e., $\varepsilon_0^a = \varepsilon_0^v = \varepsilon_0^x = 0$. With these, the simulated acceleration, speed, and position of the vehicle can be formulated as a function with respect to the error terms as follows:

$$\begin{cases} a_1^{sim} = a_1^{obs} + \varepsilon_1^a \\ v_1^{sim} = v_0^{sim} + a_1^{sim} = v_0^{obs} + a_1^{obs} + \varepsilon_1^a = v_1^{obs} + \varepsilon_1^a \\ x_1^{sim} = x_0^{sim} + v_1^{sim} = x_0^{obs} + v_1^{obs} + \varepsilon_1^a = x_1^{obs} + \varepsilon_1^a \end{cases} \quad (1)$$

$$\begin{cases} a_2^{sim} = a_2^{obs} + \varepsilon_2^a \\ v_2^{sim} = v_1^{sim} + a_2^{sim} = v_2^{obs} + \sum_{t=1}^{2} \varepsilon_t^a \\ x_2^{sim} = x_1^{sim} + v_2^{sim} = x_2^{obs} + \varepsilon_1^a + \sum_{t=1}^{2} \varepsilon_t^a \end{cases} \quad (2)$$

$$\begin{cases} a_t^{sim} = a_t^{obs} + \varepsilon_t^a \\ v_t^{sim} = v_t^{obs} + \sum_{i=1}^{t} \varepsilon_i^a \\ x_t^{sim} = x_t^{obs} + \sum_{t'=1}^{t} (t+1-t')\varepsilon_{t'}^a \end{cases} \quad (3)$$

Thus, at any time $t, t' \in \mathcal{T}$, the speed error and position error are both essentially a cumulation of the acceleration error as follows:

$$\varepsilon_t^v = \sum_{t'=1}^{t} \varepsilon_{t'}^a \quad (4)$$

$$\varepsilon_t^x = \sum_{t'=1}^{t} (t+1-t')\varepsilon_{t'}^a \quad (5)$$

Figure 2 illustrates a specific case of error propagation when $\varepsilon_5^a = 0.2$ and $\varepsilon_t^a$ at all other times are zero. It can be observed that starting from $t = 6$, the acceleration no longer reflects the influence of $\varepsilon_5^a$, while the speed and position continue to exhibit the impact of $\varepsilon_5^a$.

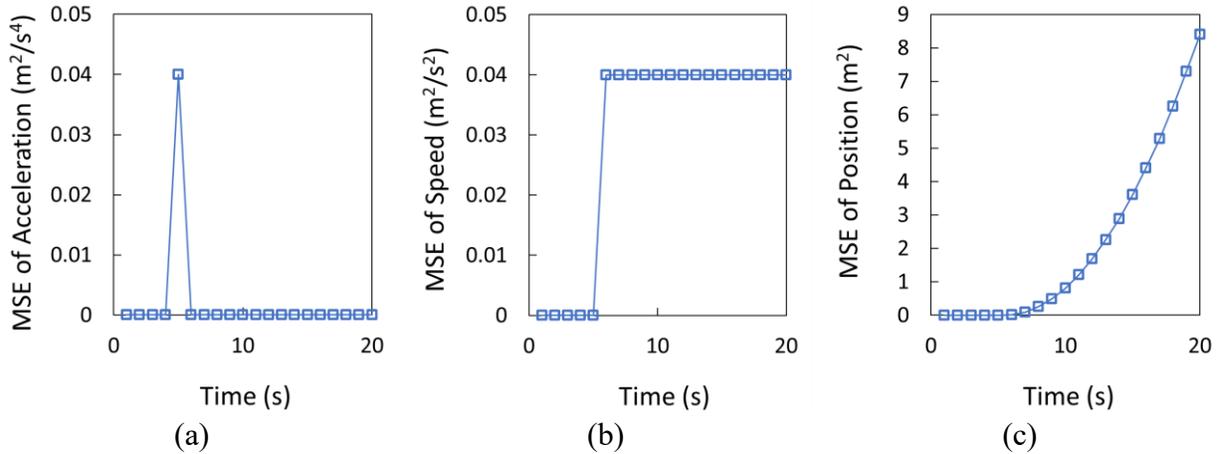

(a)          (b)          (c)

**Figure 2 Error propagation in Speed and position profiles in case of an instant error in acceleration. (a) The observed (actual) acceleration profile and simulated profiles present an error when $t = 5$, $\varepsilon_5^a = 0.2$; $\varepsilon_t^a = 0, \forall t \neq 5$; (b) The difference between observed and simulated speed; (c) The difference between observed and simulated position.**

We employ Mean Squared Error (MSE) to quantify the discrepancy between simulated outputs and real-world observations. The MSE on the acceleration and speed during the simulated time horizon are:

$$MSE^a = \frac{1}{T}\sum_{t=1}^{T}(\varepsilon_t^a)^2 \tag{6}$$

$$MSE^v = \frac{1}{T}\sum_{t=1}^{T}(\varepsilon_t^v)^2$$

$$= \frac{1}{T}\sum_{t=1}^{T}\left(\sum_{t'=1}^{t}\varepsilon_{t'}^a\right)^2$$

$$= \frac{1}{T}\left(\sum_{t=1}^{T}(T-t+1)(\varepsilon_t^a)^2 + 2\sum_{t=1}^{T-1}\sum_{t'=t+1}^{T}(T-t'+1)(\varepsilon_t^a \varepsilon_{t'}^a)\right)$$

$$= \frac{1}{T}\sum_{t=1}^{T}(\varepsilon_t^a)^2 + \frac{1}{T}\sum_{t=1}^{T}(T-t)(\varepsilon_t^a)^2 + \frac{2}{T}\sum_{t=1}^{T-1}\varepsilon_t^a \sum_{t'=t+1}^{T}(T-t'+1)(\varepsilon_{t'}^a) \tag{7}$$

Recalling the definition of $MSE^a$ and applying discrete convolution, it can be shown that:

$$MSE^v = MSE^a + \frac{1}{T}(t * (\varepsilon_t^a)^2)[T] + \frac{2}{T}\sum_{t=1}^{T-1}\varepsilon_t^a \sum_{t'=t+1}^{T}(T-t'+1)(\varepsilon_{t'}^a) \tag{8}$$

Note that $[T]$ represents the discrete convolution operation on $T$. Eq. (8) defines the relationship between the acceleration MSE and the speed MSE. The speed MSE consists of three terms: (1) the MSE on acceleration; (2) the convolution of time $T$ and acceleration error $\varepsilon_t^a$; (3) the convolution of time $T$ and errors in acceleration $\varepsilon_t^a$. Similarly, the MSE on the position is:

$$MSE^x = \frac{1}{T}\sum_{t=1}^{T}(\varepsilon_t^x)^2$$

$$= \frac{1}{T}\sum_{t=1}^{T}\left(\sum_{t'=1}^{t}(T-t'+1)\varepsilon_{t'}^a\right)^2$$

$$= \frac{1}{T}\sum_{t=1}^{T}(T+1-t)^3(\varepsilon_t^a)^2$$

$$+ \frac{2}{T}\sum_{t=1}^{T}\sum_{t'=1}^{t-1}\sum_{t''=t'+1}^{t}(t+1-t')(t+1-t'')(\varepsilon_{t'}^a\varepsilon_{t''}^a)$$

$$= \frac{(T+1)^3}{T}MSE^a + \frac{1}{T}\sum_{t=1}^{T}[-3(T+1)^2 t + 3(T+1)t^2 - t^3](\varepsilon_t^a)^2$$

$$+ \frac{2}{T}\sum_{t=1}^{T}\sum_{t'=1}^{t-1}\sum_{t''=t'+1}^{t}(t+1-t')(t+1-t'')(\varepsilon_{t'}^a\varepsilon_{t''}^a) \qquad (9)$$

With a similar analysis, Eq. (9) can be shown to be equivalent to

$$MSE^x = \frac{(T+1)^3}{T}MSE^a + \frac{1}{T}\sum_{t=1}^{T}[-3(T+1)^2 t + 3(T+1)t^2 - t^3](\varepsilon_t^a)^2 + \frac{2}{T}\sum_{t=1}^{T}(t't''\varepsilon_{t'}^a\varepsilon_{t''}^a)[t] \quad (10)$$

Eq. (10) shows the position $MSE^x$ can be formulated in terms of $\varepsilon_t^a$. Therefore, the MSE on both the speed and position of a single vehicle is influenced by the acceleration error, indicating how microscopic measurements are affected by the acceleration error.

### 2.1.2 Error Propagation of Multi-vehicle

This section extends the above analysis to a multi-vehicle scenario, aiming to illustrate the errors among the vehicles. Assuming $n \in \mathcal{N}$ vehicles and simulation time $t \in \mathcal{T}$, shown in Figure 3, given the initial state that no error on the first vehicle $n = 0$: $\varepsilon_{0,t}^a = 0, \forall t \in \mathcal{T}$, and no error on first time step $t = 0$: $\varepsilon_{n,0}^a = 0, \forall n \in \mathcal{N}$. The acceleration $a_{n,t}^{sim}$ of other vehicles at different times is determined by the car-following model $f^{CF}$, which depends on the previous state of the ego vehicle and the lead vehicle. The velocity $v_{n,t}^{sim}$ and position $x_{n,t}^{sim}$ of each vehicle are determined based on vehicle dynamics $f^D$, considering the previous velocity, acceleration, and position of the ego vehicle.

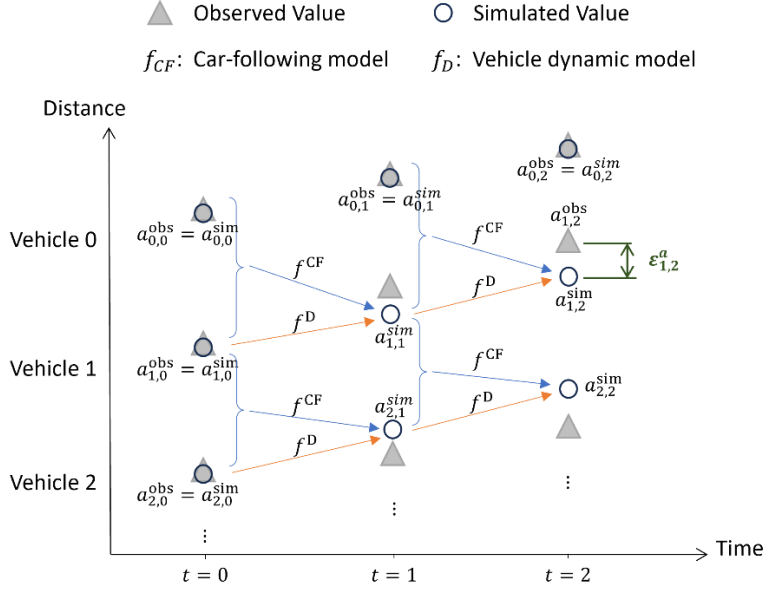

**Figure 3 Conclusion of calibration measurements adopted in the previous study.**

To enable a theoretical analysis, we consider a linear car-following model, specifically the stimulus-response model (Chandler et al., 1958), which relates the vehicle acceleration solely with the headway. However, studies (Jiang et al., 2001; Treiber et al., 2000) have found that the acceleration of a vehicle is also affected by the difference between its own speed and the speed of the preceding vehicle. Thus, we also incorporate a linear term associated with the speed difference, resulting in a linear car-following model as follows:

$$a_{n,t}^{sim} = f^{CF}\left(\Delta x_{n,t-1}^{sim}, \Delta v_{n,t-1}^{sim} | \boldsymbol{k}\right) = k_1 \Delta x_{n,t-1}^{sim} + k_2 \Delta v_{n,t-1}^{sim} + k_3 \qquad (11)$$

where $\Delta x_{n,t-1}^{sim} \coloneqq x_{n,t-1}^{sim} - x_{n-1,t-1}^{sim}$, $\Delta v_{n,t-1}^{sim} \coloneqq v_{n-1,t-1}^{sim} - v_{n,t-1}^{sim}$ and $\boldsymbol{k} = [k_1, k_2, k_3]$ are car-following model parameters.

However, the car-following model cannot replicate real-world car-following behaviors exactly, so there is an error between the simulated outcomes and real-world observations even if there are no errors in the calibrated model. We call this the model error and denote it by $r_{n,t} \coloneqq$ $a_{n,t}^{obs} - f^{CF}(\Delta x_{n,t}^{obs}, \Delta v_{n,t}^{obs}|\mathbf{k})$. With this, the observed acceleration of vehicle $n \in \mathcal{N}$ at time $t \in \mathcal{T}$ can be formulated as:

$$
\begin{aligned}
a_{n,t}^{obs} &= a_{n,t}^{sim} + \varepsilon_{n,t}^{a} \\
&= f^{CF}(\Delta x_{n,t}^{sim}, \Delta v_{n,t}^{sim}|\mathbf{k}) + \varepsilon_{n,t}^{a} \\
&= k_1(x_{n,t}^{obs} - \varepsilon_{n,t}^{x} - x_{n-1,t}^{obs} + \varepsilon_{n-1,t}^{x}) + k_2(v_{n-1,t}^{obs} - \varepsilon_{n-1,t}^{v} - v_{n,t}^{obs} + \varepsilon_{n,t}^{v}) + k_3 + \varepsilon_{n,t}^{a} \\
&= f^{CF}(\Delta x_{n,t}^{obs}, \Delta v_{n,t}^{obs}|\mathbf{k}) + k_1(-\varepsilon_{n,t}^{x} + \varepsilon_{n-1,t}^{x}) + k_2(-\varepsilon_{n-1,t}^{v} + \varepsilon_{n,t}^{v}) + \varepsilon_{n,t}^{a} \\
&= f^{CF}(\Delta x_{n,t}^{obs}, \Delta v_{n,t}^{obs}|\mathbf{k}) + r_{n,t} \quad (12)
\end{aligned}
$$

Therefore, the acceleration error at time $t$ of vehicle $n$ $\varepsilon_{n,t}^{a}$ is expressed using the speed error, position error, and model error.

$$\varepsilon_{n,t}^{a} = k_1(\varepsilon_{n-1,t-1}^{x} - \varepsilon_{n,t-1}^{x}) + k_2(\varepsilon_{n,t-1}^{v} - \varepsilon_{n-1,t-1}^{v}) + r_{n,t} \quad (13)$$

Incorporate Eqs. (4,5) into the preceding equation:

$$\varepsilon_{n,t}^{a} = \sum_{i=1}^{t-1}\left((k_1(t-i) - k_2)(\varepsilon_{n-1,i}^{a} - \varepsilon_{n,i}^{a})\right) + r_{n,t} \quad (14)$$

Given the initial state: $r_{0,t}^{a} = \varepsilon_{0,t}^{a} = 0, \forall t \in \mathcal{T}$; $r_{n,0}^{a} = \varepsilon_{n,0}^{a} = 0, \forall n \in \mathcal{N}$, the recursive formula Eq. (14) can be solved, and $\varepsilon_{n,t}^{a}$ depends solely on $r_{n,t}$.

$$\varepsilon_{n,t}^{a} = r_{n,t} + \sum_{t'=1}^{t-1}((t-t')k_1 - k_2)\left(r_{n-1,t'} - r_{n,t'} + \sum_{t''=1}^{t'-1}((t-t'')k_1 - k_2)(r_{n-2,t''} - r_{n-1,t''})\right) \quad (15)$$

This formula describes how the car-following model error $r_{n,t}$ propagates over time and through the sequence of vehicles. This propagation is a cumulative process.

Figure 4 illustrates the error propagation through the multi-vehicles in case of an instant error in acceleration: $r_{1,1}^{a} = 5$ and $r_{n,t}^{a} = 0, \forall t \neq 5$. It shows that a given error will gradually increase as it propagates through time and space.

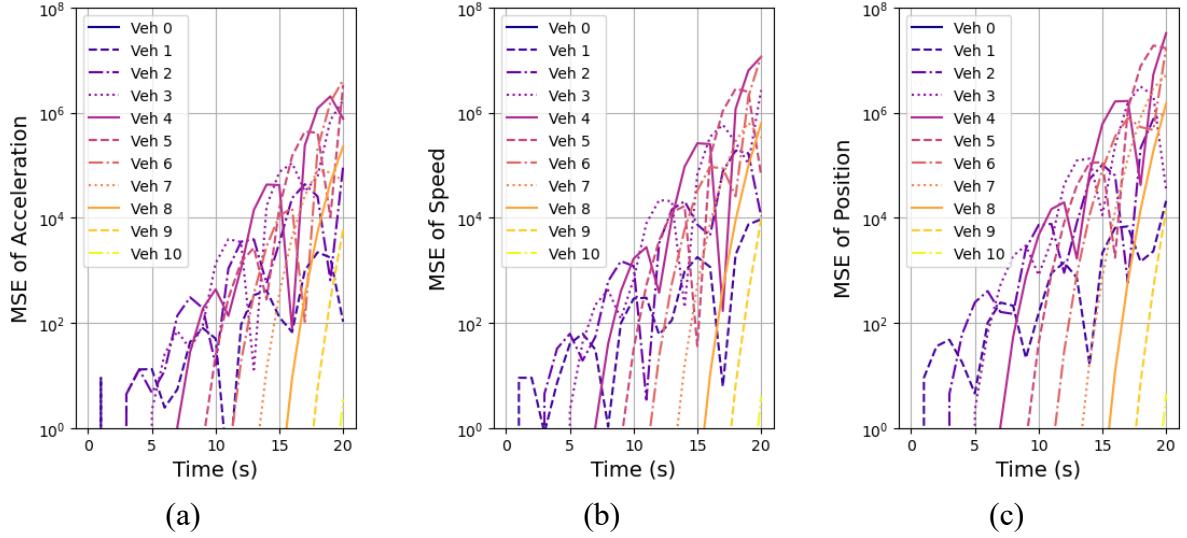

(a)                          (b)                          (c)

**Figure 4 Error propagation of multi-vehicle in speed and position profiles in case of an instant error in acceleration. (a) The error of acceleration $\varepsilon_{n,t}^a$ in case of an instant error when $t = 5, r_{1,1}^a = 5; \varepsilon_{n,t}^a = 0, \forall t \neq 1$; (b) The error of observed speed $\varepsilon_{n,t}^v$; (c) The error of observed position $\varepsilon_{n,t}^x$.**

For macroscopic measurements, it becomes necessary to compute the impact of acceleration errors, represented by model error $r_{n,i}$, on average travel time and average fuel consumption. For average travel time, the position error for vehicle n incrementally accumulates from the initial time 0 to time $t$.

$$T_{n,t_1,t_2}^{sim} = \left(x_{n,t_2}^{sim} - x_{n,t_1}^{sim}\right)/(t_2 - t_1) \tag{16}$$

$$\varepsilon_{n,t}^T = T_{n,t_1,t_2}^{sim} - T_{n,t_1,t_2}^{obs} = \left(\varepsilon_{n,t_2}^x - \varepsilon_{n,t_2}^x\right)/(t_2 - t_1) \tag{17}$$

Average fuel consumption is calculated by the VT-Micro model (Ahn et al., 2002) since it has been widely adopted in various applications. The running cost function can also be modified into other instantaneous fuel consumption methods without affecting the proposed approach.

$$e_{n,t}^{sim} = MOE_e\left(v_{n,t}^{sim}, a_{n,t}^{sim}\right) = \begin{cases} e^{\sum_{m=0}^{3}\sum_{p=0}^{3}\left(L_{m,p}^e \cdot v_{n,t}^{sim^m} \cdot a_{n,t}^{sim^p}\right)}, a_{n,t}^{sim} \geq 0 \\ e^{\sum_{m=3}^{3}\sum_{p=0}^{3}\left(M_{m,p}^e \cdot v_{n,t}^{sim^m} \cdot a_{n,t}^{sim^p}\right)}, a_{n,t}^{sim} < 0 \end{cases} \quad (18)$$

where $MOE_e\left(v_{n,t}^{sim}, a_{n,t}^{sim}\right)$ is the instantaneous fuel consumption or emission rate (L/s). $L_{m,p}^e$, $M_{m,p}^e$ are the model regression coefficients.

In this section, we have thoroughly analyzed how the microscopic and macroscopic measurements of vehicles are affected by acceleration errors. Based on these conclusions, we realize during the calibration of the car-following model, that merely focusing on either macroscopic or microscopic aspects is far from sufficient. Because the microscopic acceleration error not only affects the measurement of a single vehicle but also amplifies gradually over time and accumulates through the sequence of vehicles. Therefore, the calibration method should consider both microscopic and macroscopic traffic measurements comprehensively. In the following section, we will delve into how to consider both microscopic and macroscopic measurements in the calibration.

## 2.2 Car-following Model Calibration

Given the findings in Section 2.1, we were motivated to design a calibration approach that avoids the accumulation of localized errors. Consequently, this section proposes a bi-scale method to calibrate car-following models such that the resulting models can help reconstruct macroscopic measurements. In general, the calibration of a car-following model can be formulated as an optimization problem. Given a car-following model with an unknown parameter set $\boldsymbol{\beta}$, the objective of the calibration is to find the optimal parameter set $\boldsymbol{\beta}^*$ that minimizes the discrepancy between the observed states (for example, microscopic measurements such as position, speed, and acceleration and macroscopic measurements such as average travel time) $s_{n,t}^{obs}$ and the predicted states $s_{n,t}^{sim}$ from the model of all vehicles at all time steps:

$$\boldsymbol{\beta}^* = \arg\min_{\boldsymbol{\beta}} \frac{1}{|\mathcal{N}|}\frac{1}{|\mathcal{T}|} \sum_{n\in\mathcal{N}, t\in\mathcal{T}} \left(s_{n,t}^{obs} - s_{n,t}^{sim}\right) \tag{19}$$

where $s_{n,t}^{sim}$ is determined by the car-following model, which is typically a function of the vehicle's own state and the preceding vehicle's state at the previous time step:

$$\boldsymbol{\beta}^* = \arg\min_{\boldsymbol{\beta}} \frac{1}{|\mathcal{N}|}\frac{1}{|\mathcal{T}|} \sum_{n\in\mathcal{N}, t\in\mathcal{T}} \left(s_{n,t}^{obs} - f^{\text{CF}}\left(s_{n,t-1}^{sim}, s_{n-1,t-1}^{sim}|\boldsymbol{\beta}\right)\right) \tag{20}$$

To solve this optimization problem, the literature proposes two distinct methods: one employing microscopic measurements and the other utilizing macroscopic measurements to define the states. These approaches are discussed as follows.

**Method 1. Microscopic Calibration (MiC):** Microscopic measurements are used in microscopic calibration methods. Specifically, individual acceleration represents the most commonly employed measurement (Mo et al., 2021). The objective function of the calibration model is as follows:

$$\boldsymbol{\beta}^{\text{MiC}} = \arg\min_{\boldsymbol{\beta}} \frac{1}{|\mathcal{N}|}\frac{1}{|\mathcal{T}|} \sum_{n\in\mathcal{N}, t\in\mathcal{T}} \left(a_{n,t}^{obs} - a_{n,t}^{sim}\right)^2 \tag{21}$$

**Method 2. Macroscopic Calibration (MaC):** The macroscopic measurements include average travel time and average fuel consumption. The macroscopic measurements represent the average value of all vehicles traveling within a time period. Assuming the time $T$ is divided into $\Omega + 1$ time intervals, denoted as $\mathcal{T}_\omega$, where $\omega \coloneqq \{0,1,\ldots,\Omega\}$, $\mathcal{N}_\omega$ represents the vehicle group consisting of all vehicles that travel in the corridor during the time period $\mathcal{T}_\omega$. The calibration results for the MaC model are denoted as $\boldsymbol{\beta}^{\text{MaC}}$.

$$\boldsymbol{\beta}^{\text{MaC}} = \arg\min_{\boldsymbol{\beta}} \frac{1}{|\Omega|} \sum_{\mathcal{T}_\omega} w_1^{mac}\left(T_\omega^{obs} - T_\omega^{sim}\right)^2 + w_2^{mac}\left(e_\omega^{obs} - e_\omega^{sim}\right)^2 \tag{22}$$

where $T_\omega^{obs}$ and $e_\omega^{obs}$ are defined as the observed average travel time and average fuel consumption, respectively. These metrics represent the cumulative values for all vehicles traversing specific areas during the designated time intervals $\daleth_\omega$. The determination of specific spatial and temporal scopes should be guided by the prevailing actual conditions.

$$T_\omega^{obs} := \frac{1}{|\mathcal{N}_\omega|} \sum_{n\in\mathcal{N}_\omega, t\in\daleth_\omega} T_{n,t}^{obs} \tag{23}$$

$$e_\omega^{obs} := \frac{1}{|\mathcal{N}_\omega|} \sum_{n\in\mathcal{N}_\omega, t\in\daleth_\omega} e_{n,t}^{obs} \tag{24}$$

$T_{n,t}^{obs}$ and $T_{n,t}^{sim}$ are the actual and simulated average travel time of all vehicles during a certain time period $\daleth_\omega$, $e_{n,t}^{obs}$ and $e_{n,t}^{sim}$ are the actual and simulated average energy consumption of all vehicles during a certain time period $\daleth_\omega$.

However, these methods each consider traffic measurements from a singular perspective. The MiC is susceptible to overfitting, influenced by errors in acceleration, while the MaC falls short in capturing the intricate, micro-level behaviors of the model. To mitigate this limitation, we introduce a novel approach that integrates both microscopic and macroscopic measurements concurrently during the calibration process. This methodology enables the car-following model to accurately reflect the micro-level behaviors of vehicles, while leveraging macroscopic measurements to circumvent the risk of overfitting.

**Method 3. Bi-scale Calibration (BiC):** Using real-world trajectory and simulated trajectory, both microscopic measurements (i.e., individual acceleration) and macroscopic measurements (i.e., average travel time and average fuel consumption) are incorporated in calibration.

$$\boldsymbol{\beta}^{\text{BiC}} = \arg\min_{\beta} \frac{1}{|\varkappa||\daleth|} \sum_{n\in\varkappa, t\in\daleth} w_0^{sys}\left(a_{n,t}^{obs} - a_{n,t}^{sim}\right)^2 + \frac{1}{|\Omega|} \sum_{\daleth_\omega} w_1^{mac}\left(T_\omega^{obs} - T_\omega^{sim}\right)^2 + w_2^{mac}\left(e_\omega^{obs} - e_\omega^{sim}\right)^2 \tag{25}$$

# 3 TRAJECTORY DATA COLLECTION

To validate the proposed bi-scale calibration method, it is essential to obtain data that spans a significant temporal and spatial scale, particularly at a corridor level. Such long-duration and wide-ranging data enable a more robust validation, capturing the intricate variations and patterns inherent in car-following behaviors. Consequently, we collected 70 minutes of drone videos from a corridor and extracted the vehicle trajectories.

The video data were collected by two 4K drone cameras from 6:05–7:15 p.m. (from dusk to night) on Wednesday (November 16, 2022) over an 800 ft long segment of Park St in Madison, Wisconsin. A total of 4946 vehicle records were recorded. The period of our recording also includes the formation and dissipation of the evening peak. The bus lanes on University Ave and W Johnson St were in use at the time of recording. The studied area is shown in Figure 5. The format of the trajectory dataset is shown in Table 2.

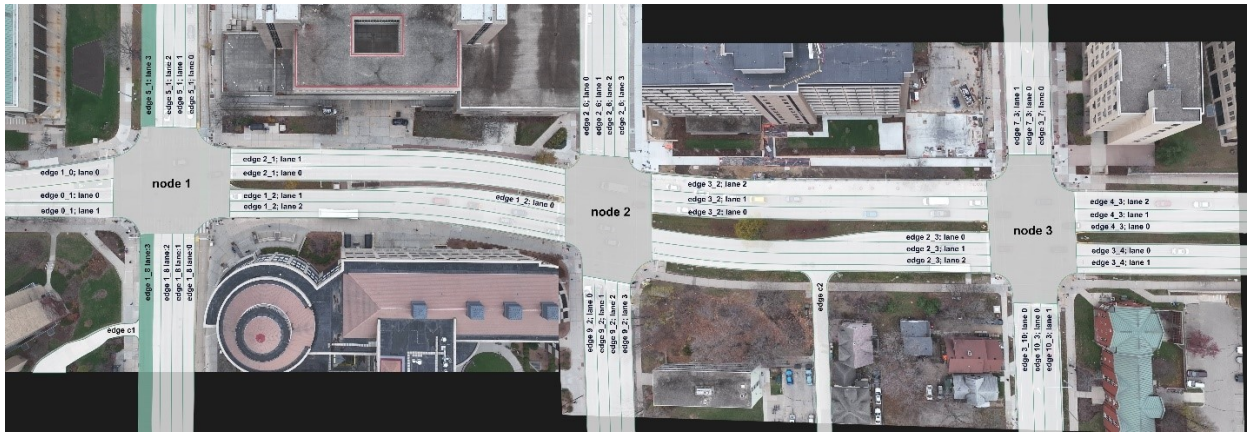

Figure 5 The study area along Park St (Source: Google Maps)

TABLE 2 Data attributes for trajectory

| Attribute | Unit | Description |
| --- | --- | --- |
| id | - | Id of vehicles. |
| time | - | Time in the format of YYYYMMDDHHMM.S. |
| x_pix | pixel | The horizontal pixel coordinate of the vehicle. |
| y_pix | pixel | The vertical pixel coordinate of the vehicle. |
| w_pix | pixel | The width of an object in pixels. |
| h_pix | pixel | The height of an object in pixels. |
| edge | - | The vehicle position of edge or intersection. |
| lane | - | The vehicle position of the lane at the given edge. If the vehicle is in an intersection, the lane is 0. |
| x_utm | m | The UTM x-coordinate of the vehicle. |
| y_utm | m | The UTM y-coordinate of vehicle. |
| t_sec | sec | The record time in seconds. |
| v | $m/s$ | Speed of the vehicle. |
| a | $m/s^2$ | Acceleration of the vehicle. |
| pre_id | - | Id of the preceding vehicle. |
| pre_v | $m/s$ | Speed of the previous vehicle. |
| delta_d | m | Distance between the outer contours of the subject vehicle and the preceding vehicle. |

The vehicle trajectory extraction and cleaning process mainly consists of two parts. The first part is vehicle trajectory extraction. With the given video taken by different drones, Step 1 stabilizes all frames by matching the feature points, Then, Step 2 merges the frames from different drones and gets the full scope of the target range. Step 3 detects and tracks the vehicle using locally trained YOLOv7 (Wang et al., 2022) and DeepSORT (Wojke et al., 2017), then get the initial trajectory data of all vehicles in pixel coordinate. The second part processes the trajectory data, Step 1 removes position offsets and then smoothes the trajectory, and Step 2 calculates the vehicle speed, acceleration, and position. In the end, the method outputs the extracted trajectory dataset.

## 3.1 Trajectory Extraction Method

This section describes how we extract vehicle trajectories from the recorded videos using drones.

### 3.1.1 Step 1: Frame Stabilize

Drifting in reference points across frames due to disturbances such as camera shake, rotation, and shifting makes finding frame-to-frame variations critical for vehicle trajectory extraction. We utilized SURF feature points detection and FLANN feature points matching algorithm to stabilize frames. Setting the reference frame for two drone camera videos, all remaining frames were matched to it. This process's error arises from two sources. First, drone position changes impacting camera angle affect feature point relationships at different altitudes. To manage this, we select feature points at similar altitudes, i.e., those on the road plane, and add a mask to the picture to cover building points. Second, the changes in lighting (sunlight, streetlights, vehicle lights) may influence the extraction of feature points. To mitigate this, we update the reference frame every 30-10 minutes to adapt to new lighting conditions. Starting from the initial reference frame, subsequent frames within 10-30 minutes are aligned to it, then the frame is updated, and the process is repeated.

### 3.1.2 Step 2: Frames Merging

To merge videos recorded by different drone cameras, we utilize the mapping relationships of feature points within each drone's field of view to merge the images. As shown in Figure 6, video frames from different cameras are merged to create a more extensive image, thereby facilitating subsequent trajectory extraction.

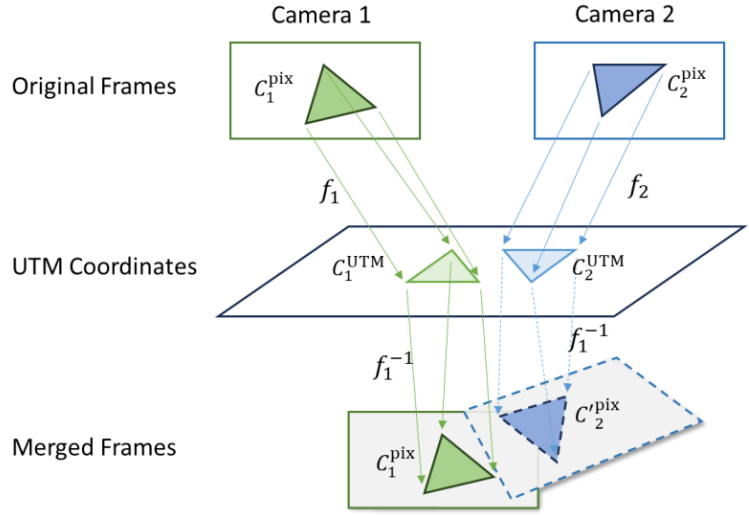

**Figure 6 Illustration of the frame merging process**

Initially, each camera requires calibration to establish the relationship between pixel coordinates of the feature points, denoted as $C_i^{pix} = [u, v]$, and the UTM coordinates of the feature points, denoted as $C_i^{UTM} = [X, Y, Z]$. The UTM coordinates are obtained from Global Positioning System (GPS) data. This relationship between these two coordinate systems is:

$$s * [u\ v\ 1]^T = K * [R|T] * [X\ Y\ Z\ 1]^T \quad (26)$$

where $K$ is the camera intrinsic matrix, $R$ is the rotation matrix, $T$ is the translation vector, and $s$ is a scaling factor.

For simplicity, we express this transformation process as follows:

$$C_i^{pix} = f_i(C_i^{utm}) = K * [R|T] * C_i^{UTM} * s^{-1}, \forall i \in D \tag{27}$$

where $C_i^{pix}$ represents the pixel coordinates of the feature points of Camera $i$. $C_i^{UTM}$ represents the actual UTM coordinates of the feature points of Camera $i$. $f_i$ represents the mapping relationship from UTM coordinates to pixel coordinates of Camera $i$.

Next, with Camera $i$ as the reference coordinate system, we calculate the pixel coordinates of the feature points within the field of view of other cameras in the pixel coordinate system of Camera $i$:

$$C'^{pix}_j = f_i^{-1}(C_j^{UTM}), \forall i, j \in D, j \neq i \tag{28}$$

where $C'^{pix}_j$ represents the pixel coordinates of the feature points within the field of view of Camera $j$, transformed into the same pixel coordinate system as Camera $i$ based on the mapping relationship from UTM coordinates to pixel coordinates of Camera $i$.

Lastly, using $C_i^{pix}$ and $C'^{pix}_j$, we transform images from other cameras through projection to overlap with the image of Camera $i$, thus accomplishing the stitching process.

In this study, we merge two video sets captured by two drones with overlapping fields of view. We manually selected 20 feature points and procured their GPS coordinates from Google Maps. Subsequently, these GPS coordinates were transformed into the Universal Transverse Mercator (UTM) coordinate system, which functioned as the world coordinate system. The chosen feature points are all situated on the ground, rendering the height difference insignificant; hence, the elevation of all feature points in the UTM coordinates is zero. Among the 20 feature points, three are within the overlapping area of Drone 1 and Drone 2's fields of view. These points were used to validate the accuracy of the transformation. The results show a negligible 0.003% discrepancy between the pixel coordinates of the three feature points in the Drone 1-pixel coordinate system and the pixel coordinates after projecting the Drone 2 image onto the Drone 1-pixel coordinate system. This minor discrepancy signifies the efficacy of our frame-merging process.

*3.1.3 Step 3: Vehicle Detection and Tracking*

To identify and track vehicles in each frame efficiently and correctly, we propose to locally train the YOLOv7 (Wang et al., 2022) and apply the model to detect vehicles in each frame of the aerial videos. To train the YOLOv7 model locally, we generate a training dataset consisting of labeled vehicle images captured from the drone video. The images are of size 640 x 640 and belong to a single class: 'veh'. Next, we apply the DeepSORT algorithm (Wojke et al., 2017) to the output of the YOLOv7 model to track the detected vehicles across frames. DeepSORT is a state-of-the-art tracking algorithm that associates detections across frames using a combination of appearance features and motion cues. Specifically, DeepSORT assigns a unique ID to each detected bounding box and tracks the movement of the bounding boxes over time. After obtaining the pixel coordinates, according to the calibration results of the reference camera in Step 2, the UTM coordinates are reversed, and the preliminary vehicle trajectory is obtained.

## 3.2 Trajectory Data Processing

*3.2.1 Step 1: Data Cleaning*

After vehicle tracking, we performed two main data-cleaning steps. The first step involved removing drifting points. Drifting points can be identified by their angular shape, and we used a vector-based approach to determine which points should be retained or removed. The angle between adjacent vectors is then computed as:

$$\theta_i = arccos\left(\frac{\vec{\gamma}_{n,t}\vec{\gamma}_{n,t-1}}{|\vec{\gamma}_{n,t}||\vec{\gamma}_{n,t-1}|}\right) \tag{29}$$

where $\vec{\gamma}_{n,t} \coloneqq (x_{n,t} - x_{n,t-1}, y_{n,t} - y_{n,t-1})$ is the direction vector.

Based on the practical considerations, points that contribute to an angular shape within 30 degrees are removed. The removed points are reconstructed using the moving average method.

The second step in the data processing procedure is track smoothing. We compared several smoothing methods and then finally chose the moving average to smooth the trajectory. we utilized a 5-point moving average to smooth the trajectory:

$$\bar{x}_t^n = (x_{t-2}^n + x_{t-1}^n + x_t^n + x_{t+1}^n + x_{t+2}^n)/5, \forall n \in \mathcal{N}, t \in \{2,3,\cdots,T-2\} \tag{30}$$

where $\bar{x}_i$ is the smoothed value at point $i$, and $x_{t-2}^n$ through $x_{t+2}^n$ are the data values within the subset used to calculate the weighted mean value. This method calculates the average value of a sliding window with a width of five data points and applies it to each data point within the trajectory.

As shown in Figure 7, the 5-point moving average method effectively reduces noise while preserving the essential features of the trajectory. However, the window size can be adjusted based on the specific characteristics of the data to achieve optimal smoothing results.

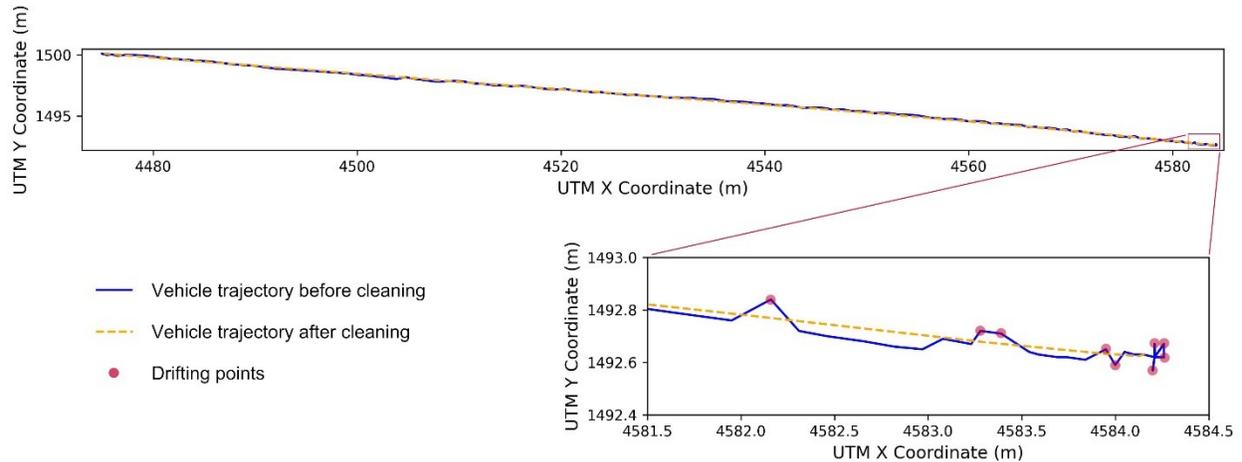

**Figure 7 Drifting points along parts of the trajectory of one vehicle.**

*3.2.2 Step 2: Determining Additional Characteristics*

To facilitate subsequent calculations, we augmented the data with vehicle speed, acceleration, and lane information. We employ filtered trajectory data to compute the vehicle's speed and acceleration, which are respectively the first and second derivatives of the position with respect to time. Vehicles are assigned to their respective lanes based on the position of their center point. Our proposed data adopt a naming convention akin to those utilized in simulation software such as SUMO and VISSIM. Intersections are denoted as 'nodes' and the connecting roads are termed 'edges.' Each 'edge' is subdivided into various 'lane'. During our data processing, the position data for each vehicle is stored under two features. The first feature refers to the 'edge' on which the vehicle is currently traveling, encompassing all lanes and intersections present on that 'edge.' The second feature specifies the exact lane occupied by the vehicle. If the vehicle is located within an intersection, the lane feature defaults to 0.

This calibrated model was subsequently employed to simulate and reconstruct vehicle trajectories. Through the comparison of multiple measurements between the original and reconstructed trajectories, we confirmed the validity of our calibrated car-following model.

## 4 EXPERIMENTS AND RESULTS

To validate the necessity of concurrently considering both macroscopic and microscopic calibration parameters in calibrating car-following models, we conducted experiments using real-world trajectory data from a corridor. This section first presents the data collection methods, vehicle trajectory data extraction, and data preprocessing. Next, details of the car-following models and the algorithm used to solve the calibration model are presented. Finally, we report and discuss the results.

The roadway includes two bus lanes (depicted in green in Fig 4). In pursuit of a more accurate vehicle simulation, vehicles are classified into two categories, 'large' and 'small,' based on their respective lengths. While these categories employ the same calibration model, the parameters within the model vary accordingly.

Our analysis incorporates two distinct types of car-following models. The first is a linear model, already explored in Section 2.1. The second includes nonlinear models, particularly the Full Velocity Difference (FVD) Model (Jiang et al., 2001) and the Intelligent Driver Model (IDM) (Treiber et al., 2000).

The FVD model, a popular car-following model, considers the full speed difference between the following and leading vehicles, making it more precise in different traffic situations.

$$a_{n,t}^{sim} = k\left[V\left(\Delta x_{n,t-1}^{sim}\right) - v_{n,t-1}^{sim}\right] + \lambda \cdot \Delta v_{n,t-1}^{sim} \tag{31}$$

$$V\left(\Delta x_{n,t-1}^{sim}\right) = \frac{V_0}{2}\left[tanh\left(\frac{\Delta x_{n,t-1}^{sim} - L_{n-1}}{b} - \beta\right) - tanh(-\beta)\right] \tag{32}$$

IDM model provides a model acceleration function as a continuous function of speed, gap, and speed difference and is expressed as follows:

$$a_{n,t}^{sim} = \bar{a}\left[1 - \left(\frac{v_{n,t-1}}{v^f}\right)^4 - \left(\frac{S(v_{n,t-1}, \Delta v_{n,t-1})}{\Delta x_{n,t-1}}\right)^2\right] \quad (33)$$

$$S(v_{n,t-1}, \Delta v_{n,t-1}) = S_0 + t_0 v_{n,t-1} - \frac{v_{n,t-1} \cdot \Delta v_{n,t-1}}{2\sqrt{\bar{a}\bar{b}}} \quad (34)$$

where $S(v_{n,t-1}, \Delta v_{n,t-1})$ is the desired space headway function and is calculated from the speed $v_{n,t-1}$ and the relative speed $\Delta v_{n,t-1}$, $v^f$ is the free flow speed, $\bar{a}$ is the maximum acceleration, $\bar{b}$ is the maximum deceleration, $t_0$ is the desired time headway, $S_0$ is the minimum space.

In the following sections, we will examine and compare the calibration results of both models, as well as discuss the volume of data necessary for their respective calibrations.

## 4.1 Results and Discussions

The estimation error of travel time and fuel consumption using three calibration methods are shown in TABLE 3.

First, among the three car-following models, the IDM model outperforms the other two car-following models. This superior performance owes to the significantly reduced average error across various measurements for the trajectories reconstructed by the IDM-based MiC, MaC, and BiC methods, in comparison to those reconstructed by the FVD model. The IDM model's superior performance, when compared to the linear model and FVD model, might be attributed to the IDM's ability to capture non-linear interactions and vehicle dynamics more effectively. The FVD model might be inherently limited due to its assumptions and simplifications, potentially making it less adaptive to varying traffic conditions.

When we focus on the performance of different calibration methods based on IDM. Among the three calibration methods, IDM-based BiC outperforms compared to IDM-based MiC and MaC. The discrepancy arises due to the fact that if we only consider travel time and fuel consumption, the increased fuel consumption caused by speed fluctuation can have a significant impact on the results. This can lead to inaccurate calibrations of maximum acceleration and maximum deceleration, in turn resulting in greater acceleration errors. Therefore, we dismiss the MaC approach and focus on the comparison between MiC and BiC. TABLE 4 presents the parameters of the car-following models calibrated using different methods. In comparison to the MiC, the BiC method resulted in a slightly lower desired speed $v^f$, a larger comfortable acceleration $\bar{a}$, smaller comfortable deceleration $\bar{b}$, an increased minimum spacing $S_0$, and a greater time headway $t_0$.

**TABLE 3 Comparison of MSE for Different Measurements using different methods.**

| CF model | Measurements | MiC | MaC | BiC |
|---|---|---|---|---|
| Linear | Acceleration ($m/s^2$) | **1.61** | 1.92 | 1.65 |
| | Speed (m/s) | 3.40 | 4.60 | **3.26** |
| | Average travel time (s) | 11.21 | 13.96 | **9.10** |
| | Average fuel consumption (L/100km) | 3.88 | 4.07 | **2.23** |
| FVD | Acceleration ($m/s^2$) | **0.69** | 1.23 | 0.73 |
| | Speed (m/s) | 1.22 | 2.19 | **1.05** |
| | Average travel time (s) | 16.15 | 7.47 | **7.26** |
| | Average fuel consumption (L/100km) | 6.86 | 4.71 | **3.55** |
| IDM | Acceleration ($m/s^2$) | **0.54** | 0.80 | 0.57 |

| | | | | | | |
|---|---|---|---|---|---|---|
| Speed (m/s) | | | 1.93 | 2.22 | **1.57** | |
| Average travel time (s) | | | 9.14 | 6.97 | **5.21** | |
| Average fuel consumption (L/100km) | | | 7.62 | 4.84 | **2.36** | |

**TABLE 4 Calibrated parameters of Car-following models using different methods.**

| CF models | Parameters | MiC | | MaC | | BiC | |
|---|---|---|---|---|---|---|---|
| | | Small vehicle | Large vehicle | Small vehicle | Large vehicle | Small vehicle | Large vehicle |
| Linear | $k_1$ | -0.053 | -0.034 | -0.562 | -0.098 | -0.078 | -0.077 |
| | $k_2$ | 0.284 | 0.145 | 2.293 | 0.162 | 0.36 | 0.183 |
| | $k_3$ | 0.918 | 0.78 | 5.778 | 0.895 | 1.042 | 0.896 |
| FVD | $k$ | 0.101 | 0.102 | 0.12 | 0.227 | 0.1 | 0.102 |
| | $\lambda$ | 0.001 | 0.031 | 0.03 | 0.034 | 0.006 | 0.011 |
| | $V_0$ (m/s) | 30.031 | 29.351 | 28.701 | 29.544 | 27.828 | 26.92 |
| | $b$ ($m/s^2$) | 10.8 | 14.732 | 10.117 | 14.864 | 14.241 | 10.358 |
| | $\beta$ | 7.736 | 7.637 | 5.323 | 9.883 | 6.283 | 8.364 |
| IDM | $v^f$ (m/s) | 22.667 | 22.137 | 19.481 | 19.546 | 17.301 | 17.137 |
| | $\bar{a}$ ($m/s^2$) | 0.494 | 0.4 | 0.528 | 0.401 | 1.256 | 1.21 |
| | $\bar{b}$ ($m/s^2$) | 2.976 | 4.123 | 3.731 | 3.724 | 3.062 | 3.023 |
| | $S_0$ (m) | 3.972 | 3.803 | 4.448 | 3.991 | 5.97 | 5.803 |
| | $t_0$ (s) | 1.127 | 1.032 | 1.593 | 0.678 | 2.261 | 2.432 |

Inaccurate parameters can further lead to skewed distributions, which in turn can result in erroneous macroscopic measurements. Figure 8 illustrates the comparison of acceleration and speed between trajectories generated by three calibration methods based on the IDM model - MiC, MaC, and BiC, and the real-world trajectory. Figure 8 (a) shows that the MiC model, calibrated using acceleration data, diverges significantly from the real-world trajectory due to its minimized maximum acceleration $\bar{a}$. All scenarios in the real-world trajectory with acceleration surpassing 0.4 are constrained in the IDM model, calibrated using the MiC method, to a maximum acceleration of $\bar{a} = 0.4$. Consequently, all scenarios with an acceleration greater than $\bar{a}$ in the original scene are weakened to less than $\bar{a}$ in the simulation.

The IDM model was calibrated using the BiC method, with a smaller desired speed $v^f$ and a larger maximum acceleration $\bar{a}$, exhibits a more similar acceleration distribution to the real-world trajectory. Therefore, due to the same reasons as the MiC model, the BiC-reconstructed trajectory is mainly distributed near $\bar{a} = 1.7$. Nevertheless, the overall error is significantly less compared to the MiC method. Moreover, the probability of acceleration fluctuation in vehicles under BiC is less than that in the real-world trajectory. This is likely because the simulation utilizes SUMO software, which embodies more rational and precise driving behavior, thereby leading to significantly less acceleration fluctuation compared to human driving.

Figure 8 (b) shows that the distribution of speed in BiC results is roughly the same as the original trajectory, while the MiC results show higher speed distributions in the high-speed range (8m/s to 12m/s) compared to the original data, mainly due to the larger desired speed $v^f$ resulting from the MiC model calibration.

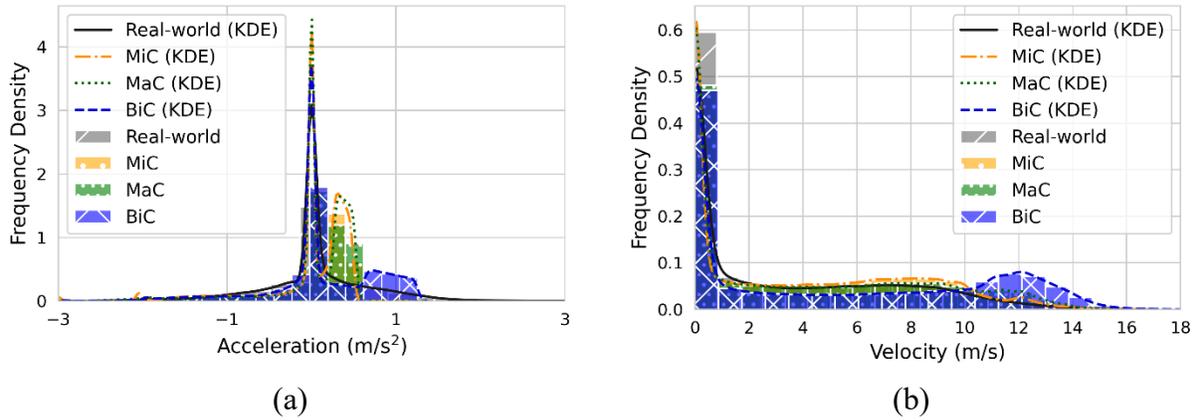

(a) (b)

**Figure 8 Comparing the distribution of microscopic measurements between real-world trajectory and reconstructed trajectory using two calibration methods.**

Taking the one vehicle on edge '3_2', lane '0' as an example, we compare the trajectory reconstruction results of the IDM model calibrated using three calibration methods, as illustrated in Figure 9. The trajectories in the figure depict the entire process of a vehicle slowing down and stopping at an intersection due to queuing, followed by restarting and accelerating when the light turns green. Among these, the MaC calibration shows the poorest results. While the accelerations predicted by both MiC and BiC seem to deviate from the same level from the real-world value, a closer examination of the actual positions and those derived from the calibrated models reveals that the position obtained from the BiC method closely matches the real position. This outcome further underscores the superiority of the BiC method, as it achieves more reasonable safety distance and desired time headway. This accuracy enables the vehicle's stop-and-go behavior near intersections to more closely resemble real-world behavior.

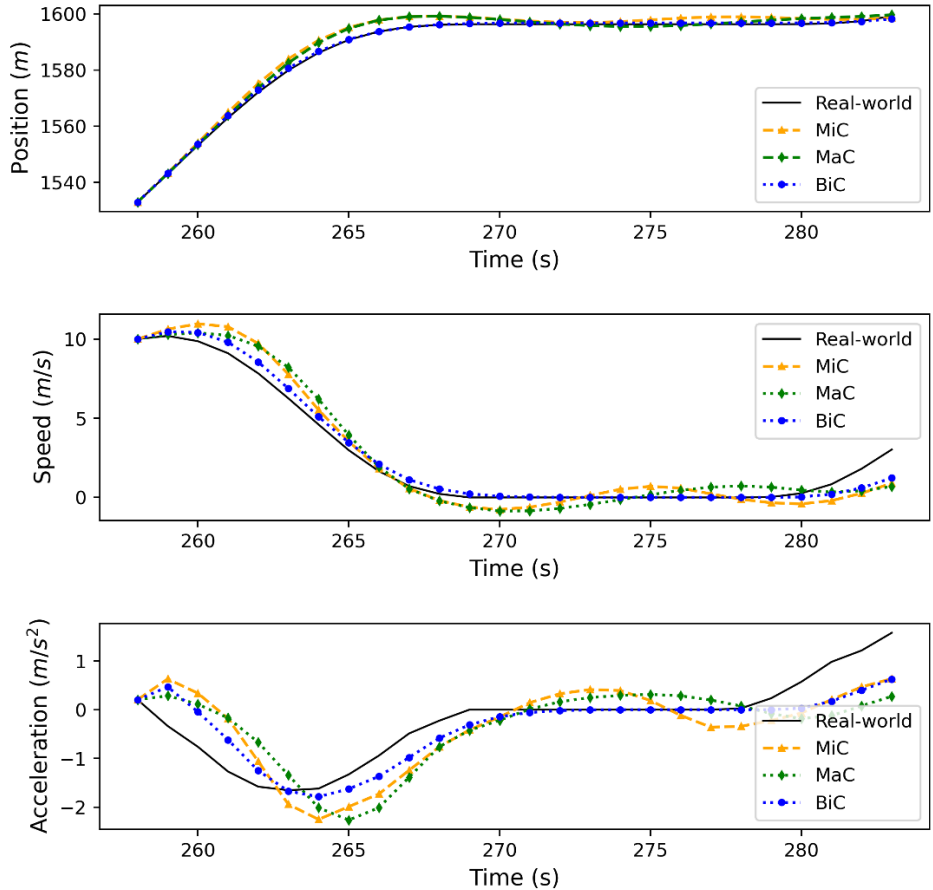

**Figure 9 Comparison of three calibrated results.**

In summary, the BiC method proves effective in offering an accurate portrayal of traffic conditions. By incorporating both microscopic and macroscopic measurements, it ensures more exact parameter estimations, enhancing the predictability for energy consumption and traffic management. In contrast, the MiC method only focuses on acceleration data, overlooking other driving dynamics. Furthermore, the inherent noise in acceleration data, particularly extreme values stemming from positional inaccuracies, poses challenges to MiC method. Without the macroscopic measurements as a regularization, filtering out these acceleration inaccuracies is tough. Thus, integrating macroscopic measurements during calibration is crucial to counteract this noise effectively.

This section mainly analysis the calibration results of model parameters and shows their impact on the distribution of microscopic measurements. As for macroscopic measurements, the comparability is impaired due to varying road lengths; hence, we conduct a specific analysis for particular road sections in the next section.

## 4.2 Results of Different Road Segments

To further analyze, we compared the real-world measurements from different segments and intersections with the estimates from various methods. In the previous section, we found that the IDM-based BiC and MiC methods perform best, and the errors from the MaC calibration were too significant to be practical. Therefore, in this section, we specifically compared the effects of the IDM-based BiC and MiC methods on different road sections.

Two representative road segments were chosen for analysis: Segment 2_1, measuring 122.2m in length, and Segment 3_2, with a length of 95.2m. An intersection, Intersection 1, spanning 32.9m in the east-west direction and 34.4m in the north-south direction, was also included. The results reveal that the best fit occurs on the road, where both micro and macro metrics closely align with the real-world trajectory. However, fitting within intersections is less accurate, particularly for acceleration and speed, representing a significant source of error.

For Segments 2_1, results are demonstrated in Figure 10. The acceleration distribution of the BiC-IDM reconstructed trajectory is more concentrated compared to the real-world trajectory, resulting in a greater proportion of vehicles being fully stationary. The acceleration of the MiC-IDM reconstructed trajectory remains aberrantly distributed around 0.4 $m/s^2$, due to incorrect acceleration calibration results. Furthermore, a higher proportion of high-speed vehicles are observed in the MiC-IDM reconstructed trajectory than in the real-world trajectory since MiC-IDM got a higher desired speed $v^f$. The mean travel time of the MiC-IDM reconstructed trajectory exceeds that of the real-world trajectory significantly, attributable to the smaller acceleration obtained during MiC-IDM calibration. This results in a longer time needed for vehicles to reach free-flow speeds, leading to higher fuel consumption than in the real-world trajectory.

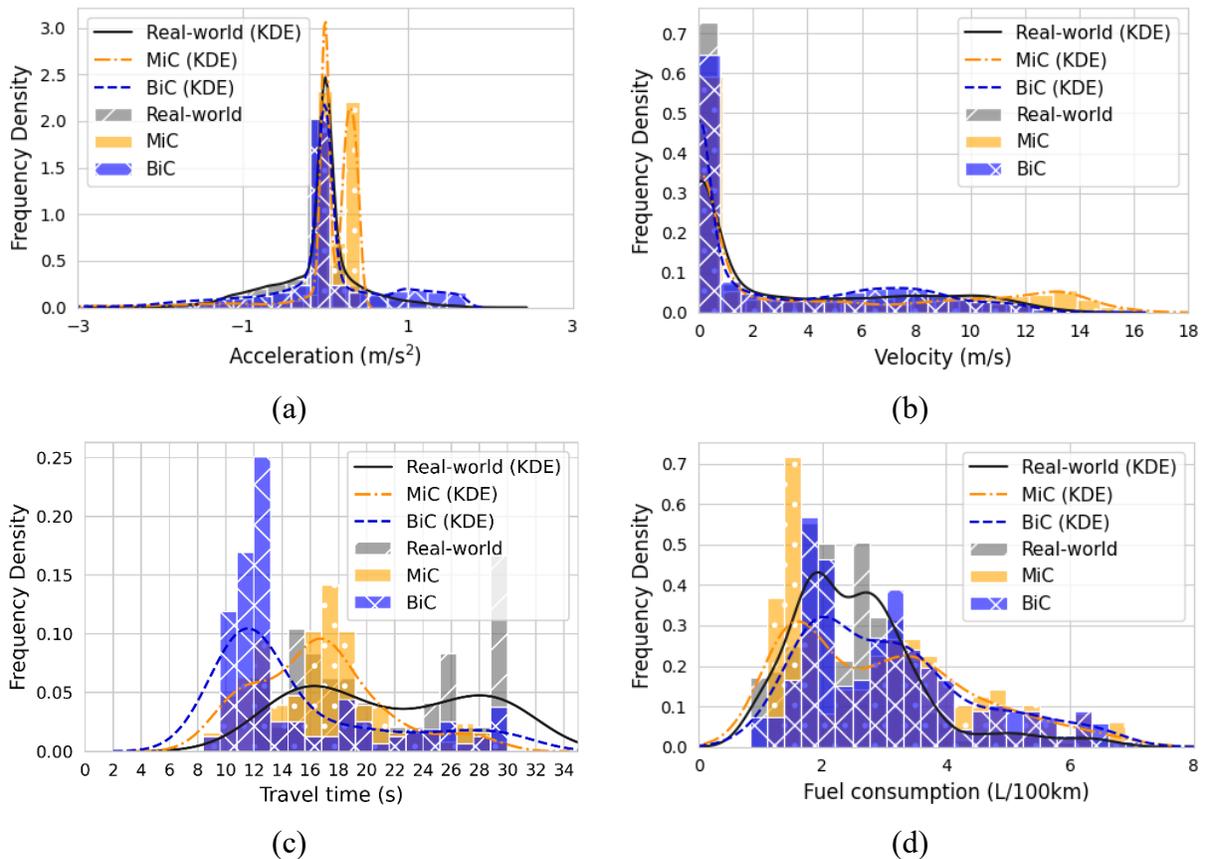

**Figure 10 Comparing the distribution of microscopic measurements between real-world trajectory and reconstructed trajectory using two calibration methods on Road 2_1.**

For Segments 3_2, results are demonstrated in Figure 11. Similar to the results of Segments 2_1, BiC-IDM reconstructed trajectory showed a more concentrated acceleration distribution than the real-world data, causing more vehicles to be stationary. The MiC-IDM reconstructed trajectory had issues with acceleration calibration, resulting in a distribution around 0.4 $m/s^2$ and a higher proportion of high-speed vehicles due to an increased desired speed. This led to higher fuel consumption in the MiC-IDM trajectory compared to the real-world data.

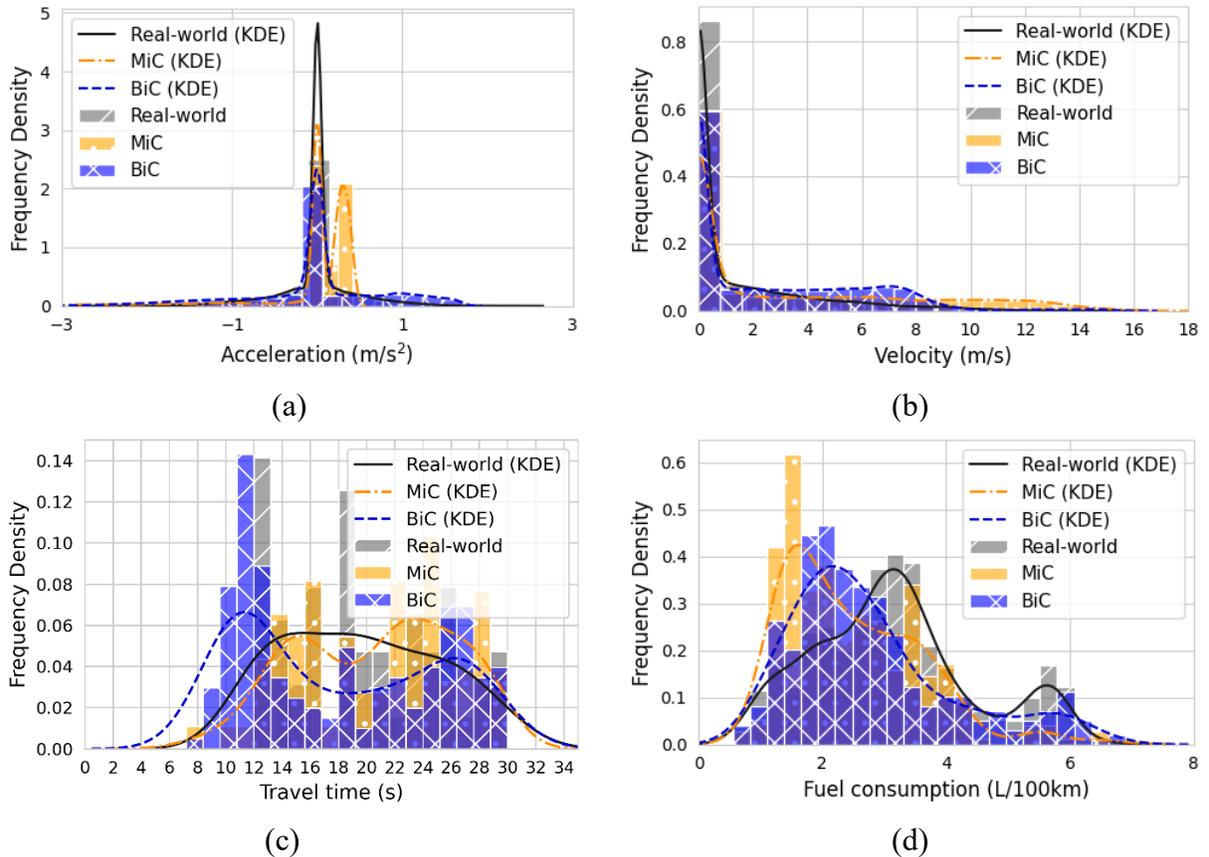

**Figure 11 Comparing the distribution of microscopic measurements between real-world trajectory and reconstructed trajectory using two calibration methods on Road 3_2.**

For internal trajectories within Intersection 1, shown in Figure 12, the disparity in acceleration and speed is noticeable due to the majority of vehicles being in acceleration mode (with 76.4% of positive accelerations in the real-world trajectory). The smaller acceleration calibrated by MiC-IDM contributes to an abnormal distribution at lower accelerations and an average speed that's skewed lower. As for travel time and fuel consumption, due to the shorter distance within the intersection, the difference between methods is not as pronounced as in the road segments.

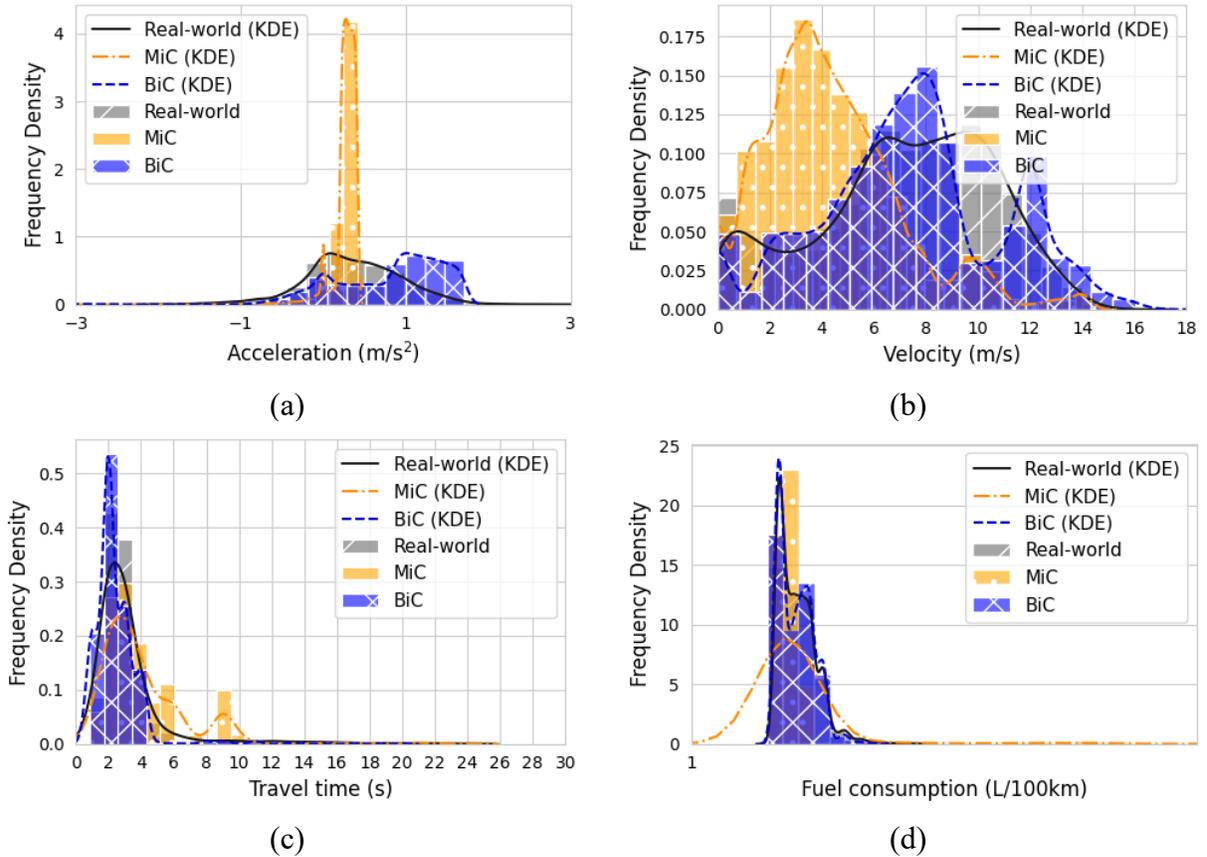

(a)　　　　　　　　　　　　　　　(b)

(c)　　　　　　　　　　　　　　　(d)

**Figure 12 Comparing the distribution of microscopic measurements between real-world trajectory and reconstructed trajectory using two calibration methods on Intersection 1.**

## 5　CONCLUSION AND FUTURE WORKS

In this study, we initiated a numerical analysis, constructing an error propagation model to determine the cumulative process of model error in speed, position, travel time, and fuel consumption across multiple vehicles involved in car-following scenarios. Our findings illustrated that macro-measurements like travel time and fuel consumption encapsulate a greater proportion of the model error from preceding vehicles and previous time periods. Therefore, integrating these macro-measurements into the calibration of vehicle models results in more robust outcomes.

Further, we leveraged real-world trajectory data from the corridor to compare three car-following model calibration methods: MiC (utilizing microscopic measurements), MaC (employing macroscopic measurements), and BiC (integrating both micro and macroscopic measurements). The results demonstrated that the IDM car-following model calibrated using the BiC method was most successful in reconstructing vehicle trajectories. The trajectories reconstructed using this method closely resembled real-world data not only in acceleration and speed distribution but also in travel time and fuel consumption.

Conversely, the traditional and common MiC method, given its exclusive focus on acceleration, tended to overfit noise in real-world trajectory accelerations. This led to an underestimation of maximum acceleration and an overestimation of desired speed, which, in the context of the experimental corridor scenario, ultimately resulted in inflated travel time and fuel consumption. Therefore, the car-following models calibrated using the MiC method are unsuitable for estimating corridor-level macro-measurements like travel time and fuel consumption.

In conclusion, our results reinforce the initial numerical analysis, underscoring the importance of incorporating both microscopic and macroscopic measurements for more robust and accurate model calibration. This research provides a solid foundation for future studies aiming to improve traffic management and energy consumption predictions by utilizing comprehensive calibration strategies.

For future work, it would be beneficial to further explore the potential of these calibration methods in different traffic scenarios and various road network structures. Moreover, a promising application direction of the proposed calibrated car-following models is analyzing the macroscopic measurements of mixed traffic flows involving connected and autonomous vehicles (CAVs) and human-driven vehicles (HVs). By performing a collaborative simulation with the car-following model and autonomous driving models, it would be possible to more accurately predict the performance of different autonomous driving behaviors (e.g., eco-drive) in terms of improving traffic efficiency and reducing energy consumption. Ultimately, these research directions may contribute to the development of more effective traffic management strategies and better support efficient and environmentally sustainable transportation systems.

# 6 ACKNOWLEDGMENTS

# 7 REFERENCES

Ahn, K., Rakha, H., Asce, M., Trani, A., Asce, M., Aerde, M. Van, 2002. Estimating Vehicle Fuel Consumption and Emissions based on Instantaneous Speed and Acceleration Levels 128, 182–190.

Alfa, A.S., Neuts, M.F., 1995. Modelling vehicular traffic using the discrete time Markovian arrival process. Transp. Sci. 29, 109–117.

Chen, C., Li, L., Hu, J., Geng, C., 2010. Calibration of MITSIM and IDM car-following model based on NGSIM trajectory datasets. Proc. 2010 IEEE Int. Conf. Veh. Electron. Safety, ICVES 2010 48–53. https://doi.org/10.1109/ICVES.2010.5550943

Hourdakis, J., Michalopoulos, P.G., Kottommannil, J., 2003. Practical procedure for calibrating microscopic traffic simulation models. Transp. Res. Rec. 1852, 130–139.

Huang, X., Sun, Jie, Sun, Jian, 2018. A car-following model considering asymmetric driving behavior based on long short-term memory neural networks. Transp. Res. Part C Emerg. Technol. 95, 346–362.

https://doi.org/10.1016/j.trc.2018.07.022

Jiang, R., Wu, Q., Zhu, Z., 2001. Full velocity difference model for a car-following theory. Phys. Rev. E 64, 17101. https://doi.org/10.1103/PhysRevE.64.017101

Kesting, A., Treiber, M., 2008. Calibrating car-following models by using trajectory data methodological study. Transp. Res. Rec. 148–156. https://doi.org/10.3141/2088-16

Li, L., Chen, X.M., Zhang, L., 2016. A global optimization algorithm for trajectory data based car-following model calibration. Transp. Res. Part C Emerg. Technol. 68, 311–332. https://doi.org/10.1016/j.trc.2016.04.011

Ma, L., Qu, S., 2020. A sequence to sequence learning based car-following model for multi-step predictions considering reaction delay. Transp. Res. Part C Emerg. Technol. 120, 102785. https://doi.org/10.1016/j.trc.2020.102785

Mo, Z., Shi, R., Di, X., 2021. A physics-informed deep learning paradigm for car-following models. Transp. Res. Part C Emerg. Technol. 130, 103240. https://doi.org/10.1016/j.trc.2021.103240

Papathanasopoulou, V., Antoniou, C., 2015. Towards data-driven car-following models. Transp. Res. Part C Emerg. Technol. 55, 496–509. https://doi.org/10.1016/j.trc.2015.02.016

Pourabdollah, M., Bjarkvik, E., Furer, F., Lindenberg, B., Burgdorf, K., 2018. Calibration and evaluation of car following models using real-world driving data. IEEE Conf. Intell. Transp. Syst. Proceedings, ITSC 2018-March, 1–6. https://doi.org/10.1109/ITSC.2017.8317836

Punzo, V., Borzacchiello, M.T., Ciuffo, B., 2011. On the assessment of vehicle trajectory data accuracy and application to the Next Generation SIMulation (NGSIM) program data. Transp. Res. Part C Emerg. Technol. 19, 1243–1262.

Punzo, V., Ciuffo, B., Montanino, M., 2012. Can results of car-following model calibration based on trajectory data be trusted? Transp. Res. Rec. 11–24. https://doi.org/10.3141/2315-02

Punzo, V., Simonelli, F., 2005. Analysis and comparison of microscopic traffic flow models with real traffic microscopic data. Transp. Res. Rec. 53–63. https://doi.org/10.3141/1934-06

Shang, M., Rosenblad, B., Stern, R., 2022. A Novel Asymmetric Car Following Model for Driver-Assist

Enabled Vehicle Dynamics. IEEE Trans. Intell. Transp. Syst. 23, 15696–15706. https://doi.org/10.1109/TITS.2022.3145292

Song, G., Yu, L., Geng, Z., 2015. Optimization of Wiedemann and Fritzsche car-following models for emission estimation. Transp. Res. Part D Transp. Environ. 34, 318–329. https://doi.org/10.1016/j.trd.2014.11.023

Song, G., Yu, L., Xu, L., 2013. Comparative analysis of car-following models for emissions estimation. Transp. Res. Rec. 12–22. https://doi.org/10.3141/2341-02

Treiber, M., Hennecke, A., Helbing, D., 2000. Congested traffic states in empirical observations and microscopic simulations. Phys. Rev. E - Stat. Physics, Plasmas, Fluids, Relat. Interdiscip. Top. 62, 1805–1824. https://doi.org/10.1103/PhysRevE.62.1805

Treiber, M., Kesting, A., 2013. Microscopic Calibration and Validation of Car-Following Models – A Systematic Approach. Procedia - Soc. Behav. Sci. 80, 922–939. https://doi.org/10.1016/j.sbspro.2013.05.050

Vasconcelos, L., Neto, L., Santos, S., Silva, A.B., Seco, Á., 2014. Calibration of the gipps car-following model using trajectory data. Transp. Res. Procedia 3, 952–961. https://doi.org/10.1016/j.trpro.2014.10.075

Wang, C.-Y., Bochkovskiy, A., Liao, H.-Y.M., 2022. {YOLOv7}: Trainable bag-of-freebies sets new state-of-the-art for real-time object detectors. arXiv Prepr. arXiv2207.02696.

Wojke, N., Bewley, A., Paulus, D., 2017. Simple Online and Realtime Tracking with a Deep Association Metric, in: 2017 IEEE International Conference on Image Processing (ICIP). pp. 3645–3649. https://doi.org/10.1109/ICIP.2017.8296962

Zhu, M., Wang, X., Wang, Y., 2018. Human-like autonomous car-following model with deep reinforcement learning. Transp. Res. Part C Emerg. Technol. 97, 348–368. https://doi.org/10.1016/j.trc.2018.10.024